# PRECISION MEASUREMENT OF THE PROTON CHARGE RADIUS IN ELECTRON PROTON SCATTERING.

(PROJECT OF EXPERIMENT)

## A. Vorobyev

*Petersburg Nuclear Physics Institute,*
*NRC "Kurchatov Institute", Gatchina, Russia*

e-mail: vorobyov_aa@pnpi.nrcki.ru

April 22, 2019

**Abstract**. This report presents a project of an experiment for precision studies of the elastic electron proton scattering in the low momentum transfer region. The project is based on an innovative experimental method which allows for detection of recoil protons in a hydrogen "active target". The goal of the experiment is to measure the *ep* elastic scattering differential cross section in the $Q^2$ range from 0.001 GeV$^2$ to 0.04 GeV$^2$ with 0.1% relative and 0.2% absolute precision and to determine the proton charge radius with a sub-percent precision. The important advantages of the proposed experiment are low radiative corrections, inherent to the recoil proton method, and absolute measurements of the differential cross sections. The experiment will be performed in the 720 MeV electron beam of the Mainz electron accelerator MAMI. This accelerator can provide an electron beam with optimal for this experiment parameters. The Proposal was approved by the MAMI Program Advisory Committee, and a special Agreement aimed at realization of this experiment was signed between Petersburg Nuclear Physics Institute and Institute of Nuclear Physics of the Mainz University. Also, scientists from the following institutes will participate in this project: GSI Centre for Heavy Ion Research, Germany; Joint Institute for Nuclear Research, Russia; William and Mary College, USA; Mount Allison University, Canada; University of Regina, Canada; Saint Mary's University, Canada. The first measurements should be started in 2020.

## Introduction

This project is motivated by the striking discrepancy between precise measurements of the proton charge radius $R_p = \langle r_p^2 \rangle^{1/2}$ in the muonic hydrogen Lamb shift experiments performed at PSI by the CREMA Collaboration ( $R_p$ = 0.84087(39) fm [1]) and the radius determined in the electron-proton elastic scattering experiments: $R_p$ = 0.879(5)$_{stat}$(6)$_{syst}$ fm, A1 collaboration at Mainz [2], and $R_p$ = 0.875(10) fm, Thomas Jefferson National Accelerator Facility [3]. The so-called "proton radius puzzle" is widely discussed in the scientific community. Various reasons for the observed discrepancy are under discussion, including possible contribution of some exotic particle coupling differently to electrons and muons. One should note that the experimental data in the low $Q^2$ region obtained in the previous *ep* scattering experiments are not sufficient for precision extraction of the proton radius. Therefore, extrapolation from higher $Q^2$ regions is used, and the result becomes dependent on the $Q^2$ shape of the proton form factors. Another problem might be related to application of the radiative corrections to the measured differential cross sections.

In all previous *ep* scattering experiments, the differential cross sections were determined by measuring the angular distribution of the scattered electrons selected by momentum with the magnetic spectrometers. In this case, the radiative corrections are quite large (~10%), strongly depending on the selection procedure of the scattered electrons. These corrections are $Q^2$ dependent, and they may influence the extracted value of the proton radius. In principle, the level of the introduced radiative corrections could be controlled by the absolute measurements of the differential cross sections. However, no such measurements exist until today. In this context, new high precision studies of the *ep* elastic scattering in the low $Q^2$ region, including absolute measurements of the differential cross sections, are desired.

The first new generation *ep* scattering experiment in the low $Q^2$ region is the PRad experiment at Jlab [4]. In this experiment, the electron scattering on a hydrogen gas jet-like target is studied in the $Q^2$ range from $2 \cdot 10^{-4}$ GeV$^2$ to $8 \cdot 10^{-2}$ GeV$^2$. The angle and energy of the scattered electron are detected with a forward tracker and a forward calorimeter. The estimated radiative corrections are at the level of 15%. The elastic *ep* scattering cross section is normalized to the simultaneously measured Møller scattering cross section. The PRad experiment started taking data in 2016.

The main feature of the experiment to be realized in our project is detection of the recoil protons with a specially designed hydrogen "active target". The goal is to measure the *ep* elastic scattering differential cross section in the $Q^2$ range from 0.001 GeV$^2$ to 0.04 GeV$^2$ with 0.1% relative and 0.2% absolute precision and to determine the proton radius with a sub-percent precision. The important advantages of the proposed experiment are low radiative corrections, inherent to the recoil proton method, and absolute measurements of the differential cross sections. The experiment will be performed in the 720 MeV electron beam at the Mainz electron accelerator MAMI. This accelerator can provide an electron beam with practically optimal for this experiment parameters, as it was demonstrated in a special test run in September 2017. The Proposal was approved by the MAMI Program Advisory Committee, and a special Agreement aimed at realization of this experiment was signed between PNPI and INP Mainz.

**1. Experimental overview**

In the leading order approximation, the *ep* elastic scattering differential cross section at high electron energies is given by the following expression:

$$\frac{d\sigma}{dt} = \frac{\pi\alpha^2}{t^2} \left\{ G_E^2 \left[ \frac{(4M + t/\varepsilon_e)^2}{4M^2 - t} + \frac{t}{\varepsilon_e^2} \right] - \frac{t}{4M^2} G_M^2 \left[ \frac{(4M + t/\varepsilon_e)^2}{4M^2 - t} - \frac{t}{\varepsilon_e^2} \right] \right\}, \quad (1)$$

where $\alpha = 1/137$, $\varepsilon_e$ – the initial total electron energy, $M$ – the proton mass, $t = 2MT_R = -Q^2$, $T_R$ – the recoil proton energy, $G_E$ – the electric form factor, $G_M$ – the magnetic form factor. At low $Q^2$, the form factors $G_E$ and $G_M$ can be represented by the power series expansions:

$$G_{E,M}(Q^2)/G_{E,M}(0) = 1 - <r_p^2> Q^2 / C_2 + <r_p^4> Q^4 / C_4 - <r_p^6> Q^6 / C_6 + ... \quad . \qquad (2)$$

Here $C_n = (n+1)!$ and $r_p$ stands for $r_{pE}$ or $r_{pM}$ – the proton charge and magnetic radii, respectively. An example of $d\sigma/dt$ in the small $Q^2$ region is shown in Fig. 1.

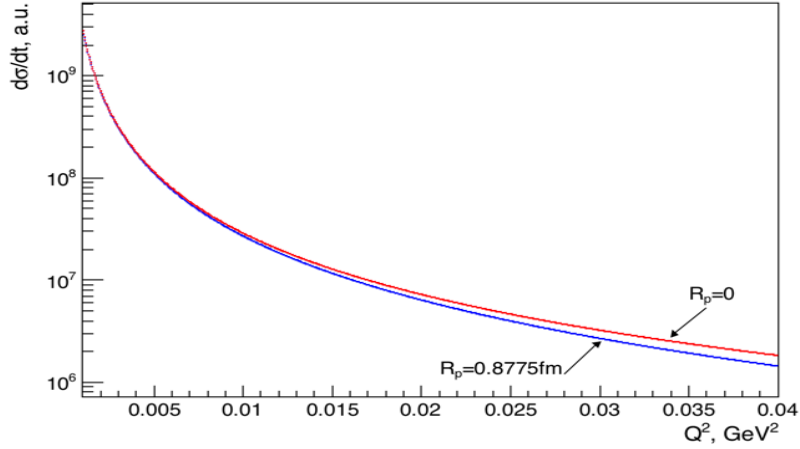

**Fig. 1.** Differential cross section for *ep* elastic scattering at $\varepsilon_e = 720$ MeV calculated for point-like proton and for $R_p = 0.8775$ fm with the proton form factor taken from [2]

The differential cross sections will be measured for $\varepsilon_e = 720$ MeV in the $Q^2$ range from 0.001 GeV$^2$ to 0.04 GeV$^2$. The sensitivity of $d\sigma/dt$ to the proton radius in this $Q^2$ range is rather low, as it is demonstrated in Fig.1. The cross sections corresponding to $R_p = 0.88$ fm and $R_p = 0.84$ fm differ only by 1.3% at $Q^2 = 0.02$ GeV$^2$. That means that at least 0.2% precision in measurements of $d\sigma/dt$ is needed to distinguish reliably between these two options. This experiment is designed to reach such precision.

The experimental method is based on detection of the recoil protons in a specially designed hydrogen Time Projection Chamber (hydrogen TPC) operating as an "active target" in combination with a high precision Forward Tracker (FT) for detecting the scattered electrons. The hydrogen TPC has been first developed at PNPI. It was used in several experiments including experiments WA9 and NA8 at CERN for studies of small-angle $\pi p$ and $pp$ scattering at high energies [5]. A new advanced version of the high-pressure hydrogen TPC will be used in this experiment. Figure 2 shows a schematic view of the proposed experimental setup. It consists of a hydrogen TPC, a Multiwire proportional chamber (MWPC) based Forward Tracker, and a beam monitoring system.

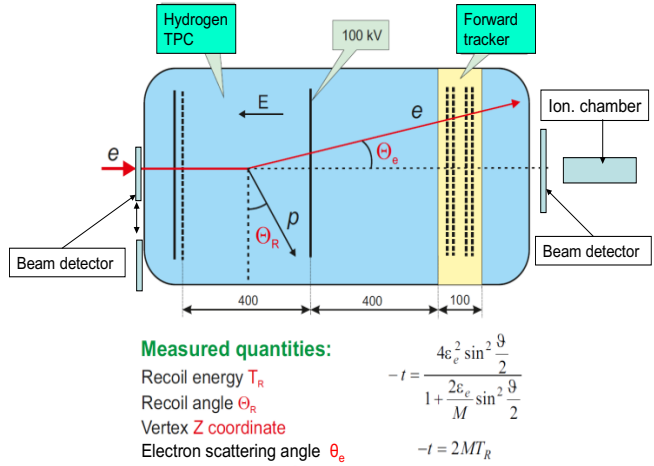

Fig. 2. Schematic view of the combined TPC & FT detectors

The hydrogen TPC operates in the ionization mode (no gas amplification). It allows to measure the recoil proton energy $T_R$, the recoil proton angle $\theta_R$, and the Z-coordinate (along the beam direction) of the vertex $Z_V$. The Cathode-Grid distance (drift space) is 400.0 mm. The electron total drift time is about 100 μs. The electron drift velocity will be measured with 0.01% precision directly in this experiment in a special mode. The outer diameter of the anode is 600 mm. The TPC will operate at the gas pressure up to 20 bar with the maximal energy of the recoil protons stopped in the TPC sensitive volume about 10 MeV. For higher proton energies, the TPC measures the energy deposited in the sensitive volume (for example, 5 MeV for 20 MeV protons). Also, there is a possibility to use the $CH_4$ gas filling. In this case the maximal energy of the protons stopped inside the TPC is 22 MeV. The $H_2$ pressure and temperature in the TPC will be controlled with 0.01% precision. As the result, the target thickness (the number of protons in the selected TPC volume) will be known with 0.04% precision, without any «wall effects».

The forward tracker (FT) is designed for high absolute precision in measuring the X- and Y- coordinates of the electron track relative to the beam line. Together with the Z-coordinate of the recoil proton, measured by the TPC in combination with the track timing, this provides a precision measurement of the electron scattering angle $\theta_e$. Also, the FT measures the arrival times of the scattered electrons. The FT acceptance is from 20 mrad to 460 mrad with detection efficiency close to 100%. The FT contains four planes of MWPCs with the cathode strip readout. The key element of the MWPC is the cathode plane with orthogonal cathode-to-anode wire strips. It provides the absolute measurement of the track coordinate along the anode wire. Using the centre-of-gravity method, the coordinate of each detected track is determined with ~30 μm precision. The MWPC strip plane is fabricated in such a way that it guarantees the absolute linear scale with ~0.02% precision.

The FT and TPC are assembled in a common vessel in separated volumes. The TPC operates with ultra-clean hydrogen at up to 20 bar gas pressure, while an Ar + 2%CH$_4$ gas mixture is used in the FT at the gas pressure equalized with that in the TPC. The gas purity, pressure, and temperature will be maintained with the help of the specially designed gas circulation/purification systems.

The beam detectors have several functions: tracing the beam line and controlling its stability; measuring the arrival times of the beam electrons; absolute counting of the beam electrons for determination of the absolute cross section. The first of these functions is provided by the pixel detectors. The second and the third functions will be realized with the fast scintillator detectors (SC) downstream of the TPC&FT detector. Also, a high-pressure ionization chamber will be used for the beam current control and for evaluation of the pile-up correction in the absolute counting of the beam rate with the SC detectors (the beam rate will be around 2 MHz). With this system, the integrated beam rate will be measured with 0.05% absolute precision. The upstream beam detectors (the SC detectors and pixels) will be used only for calibration purposes, and they will be in the out-of-beam position during the physics runs in order to reduce the amount of material in the beam line. The beam enters the TPC through a 400 μm Be window. The parameters of the electron beam to be used in this experiment are presented in Table 1. It should be pointed out that these parameters are practically ideal for such measurements.

**Table 1**
Parameters of the electron beam designed for this experiment

| Beam energy | 720 MeV |
|---|---|
| Beam energy resolution | < 20 keV (1σ) |
| Absolute beam energy precision | ±150 keV (1σ) |
| Beam intensity (main run) | 2·10$^6$ e/s |
| Beam intensity for calibration | 10$^4$ e/s and 10$^3$ e/s |
| Beam divergency | ≤ 0.5 mrad (1σ) |
| Beam size | ≤ 0.2 mm (1σ) |
| Duty factor | 100% |

## 2. Measurement procedure

*Transfer momentum determination.* As it follows from Eq.(1), the *ep* elastic scattering differential cross section is determined by the transfer momentum, and it is practically independent of the electron energy at $\varepsilon_e \geq 500$ MeV in the considered low $Q^2$ region. The transfer momentum $-t = Q^2$ can be found either from the recoil proton energy $T_R$ or from the electron scattering angle $\theta_e$. The advantage of the $T_R$ method is the determination of the transfer momentum independently of the incident electron energy $\varepsilon_e$: $-t = 2MT_R$. Therefore, measuring the transfer momentum by the $T_R$ method, we avoid the influence of the beam energy losses before the *ep* collision on the measured $d\sigma/dt$. This is especially important for the *ep* scattering because of the radiation losses of the electrons in the materials upstream of the *ep* collision point. In contrast, the transfer momentum determined *via* the electron scattering angle $\theta_e$ depends on $\varepsilon_e$. A possible tail in $\varepsilon_e$ results in a tail in the measured $Q^2$ distribution and thus disturbs the $d\sigma/dt$ measurement. On the other hand, the $\theta_e$ scale can be determined with high absolute precision. This allows to perform precise $T_R$ scale calibration using the measured $\theta_e - T_R$ correlation plot (Fig.3 left panel). We call this procedure a "self-calibration of the $T_R$ scale" as it does not require any special measurements and can be performed using the full set of the collected experimental data. This is an essential point of this experimental method. The electron scattering angle is measured with 0.02% absolute precision. Furthermore, the absolute energy of the 720 MeV electron beam is known with 0.02% precision. This allows to calibrate the $T_R$ scale with 0.04% relative and 0.08% absolute precision.

*Radiative corrections.* An important advantage of the recoil method is relatively small radiative correction to the measured value of $d\sigma/dt$. Figure 3 (right panel) shows the main diagrams of the radiative processes in the *ep* scattering. In the previous experiments, where the transfer momentum was determined by measuring the angle and the momentum of the scattered electron, the main contributions to the radiative corrections came from the real and virtual gamma emission by the electron (diagrams v2, r1, and r2). On the contrary, they should cancel each other almost exactly when the transfer momentum is determined by the recoil energy $T_R$, as it is described in a recent publication [6]. In fact, this cancellation is under condition that there are no cuts introduced in the scattered electron distributions. Figure 3 (left panel) demonstrates the radiative tail in the $\theta_e$ distribution corresponding to a selected recoil energy about $T_R = 5$ MeV. This tail falls into acceptance of the Forward Detector. In the analysis it can be included into the number of the elastic scattering events corresponding to the selected $\Delta T_R$, thus minimizing the effect of cuts in the $\theta_e$ distributions.

*Elimination of background.* The main background in this experiment is expected from the inelastic scattering reaction $ep \rightarrow ep\pi^0$. The $T_R - \theta_e$, $T_R - \theta_R$, and $\theta_R - \theta_e$ correlations can be used to eliminate this background. As it was demonstrated with the dedicated Monte-Carlo simulations, these correlations allow to separate the elastic scattering from the background without noticeable losses of the elastic scattering events.

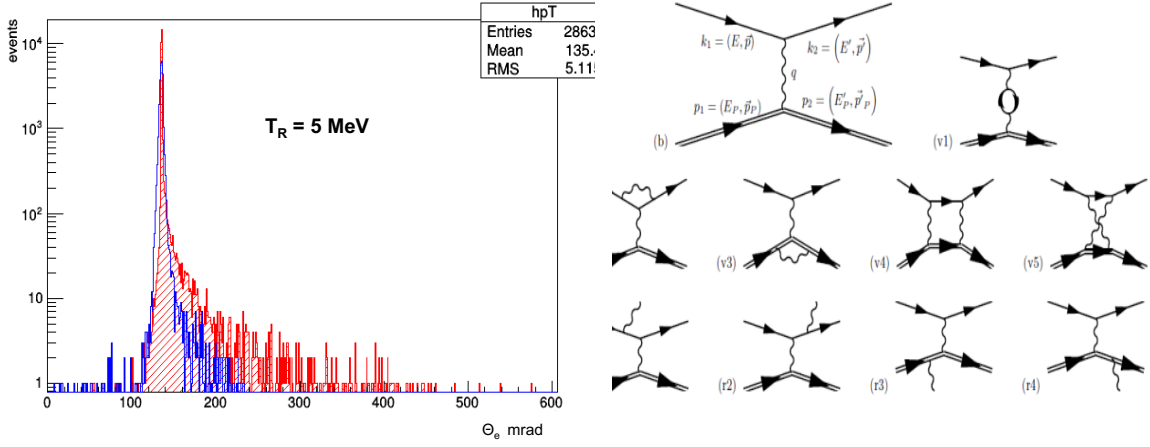

**Fig. 3.** Right panel: angular distribution of the 720 MeV electrons after *ep* collisions calculated with a MC generator taking into account all radiative corrections (red colour). This distribution corresponds to the selected recoil energy $T_R$ = [4.9, 5.1] MeV. For comparison, the angular distribution of the electrons due to multiple Coulomb scattering is shown in blue colour. The peak position in this distribution is used to calibrate the $T_R$ scale. Left panel: main diagrams for radiative processes in *ep* elastic scattering.

*Trigger and acquisition.* An adjustable combination of signals from the TPC anodes exceeding some threshold values will be used for triggering the readout system (the TPC self-triggering mode). This is the most safe and effective triggering option. The expected trigger rate will be around 50 Hz. The acquisition system will provide continuous data flow without introducing any dead time. After receiving a trigger signal, the information from all detectors which appeared in a 100 μs time interval before the arrival of the trigger is readout from the pipeline and sent to DAQ. The efficiency in detection of the *ep* events triggered by the TPC should be close to 100% in the measured $Q^2$-range.

*Selection of true ep collisions.* The trigger is a recoil signal ($T_R \geq 0.3$ MeV) detected in the TPC at the time $t_R$. The maximum drift time in the TPC is 100 μs. Therefore, any beam electron appearing in the TPC in the time interval $t_R - 100$ μs $\leq t \leq t_R$ should be considered as a candidate for the recoil parent particle. The electron beam intensity is $2 \cdot 10^6$ electrons per second. This means that the average number of the *ep* candidates at this stage is 200. However, by requiring a signal in the FT in the searched 100 μs time interval (rejection factor ~ 40) and applying back tracing of the electron trajectory determined by the FT planes (rejection factor ~ 10), one can reduce the number of false candidates to less than one per a true *ep* event. In addition, the $\theta_e - T_R$ correlation provides a powerful background rejection. Together with the previous selection steps, this allows to select the true *ep* events on a high confidence level with ~100% detection efficiency.

## 3. Statistical and systematic errors

Table 2 presents a summary of various sources of systematic errors contributing to the systematic error in the measured differential cross section of *ep* elastic scattering.

**Table 2.** Contribution of various sources to the systematic error in the measured $d\sigma/dt$

|    |                                           | Syst.error % | comments |
|----|-------------------------------------------|--------------|----------|
| 1  | Drift velocity, $W$                       | 0.01         |          |
| 2  | High Voltage, HV                          | 0.01         |          |
| 3  | Temperature, $K$                          | 0.015        |          |
| 4  | Pressure, $P$                             | 0.01         |          |
| 5  | $H_2$ density, $\rho_p$                   | 0.025        | Linear sum of errors 3 and 4 |
| 6  | Target length, $L_{tag}$                  | 0.02         |          |
| 7  | Number of protons in target, $N_p$        | 0.045        | Linear sum of errors 5 and 6 |
| 8  | Number of beam electrons, $N_e$           | 0.05         | Beam detector counts |
| 9  | Detection efficiency of $ep$ events       | 0.05         |          |
| 10 | Electron beam energy, $\varepsilon_e$     | 0.02         |          |
| 11 | Electron scattering angle, $\theta_e$     | 0.02         |          |
| 12 | $t$-scale calibration, $T_R$ relative     | 0.04         | Twice error 11 |
| 13 | $t$-scale calibration, $T_R$ absolute     | 0.08         | Linear sum of errors 10 and 11 |
|    | **$d\sigma/dt$, relative**                | 0.1          | Twice error 12 plus error 9 |
|    | **$d\sigma/dt$, absolute**                | 0.2          | Twice error 13 plus errors 7, 8, 9 |

As it follows from Table 2, the expected systematic errors in the measured $d\sigma/dt$ are 0.1% and 0.2% in relative and absolute measurements, respectively. In order to estimate the level of statistical and systematic errors in the procedure of extraction of the proton radius from the measured differential cross section, we simulated such cross section using the proton form factor available from analysis of the existing experimental data [7]. About 70 million of the $ep$ elastic scattering events were generated in the $Q^2$ range from 0.001 GeV$^2$ to 0.04 GeV$^2$, that corresponds to the statistics to be collected in our experiment during 45 days of data taking. Then the proton radius was extracted from fitting the simulated pseudo-data with the cross section calculated using a power series expansion of the proton electric form factor up to the $Q^8$ term:

$$G_E(Q^2) = A(1 - \langle r_p^2 \rangle B_2 \cdot Q^2/C_2 + \langle r_p^4 \rangle \cdot B_4 \cdot Q^4/C_4 - \langle r_p^6 \rangle \cdot B_6 \cdot Q^6/C_6 + \langle r_p^8 \rangle \cdot B_8 \cdot Q^8/C_8), \qquad (3)$$

where $B_n = (5.068)^n$, $C_n = (n+1)!$, $n = 2,4,6,8;$ $\langle r_p^n \rangle$ and $Q^n$ are expressed in fm$^n$ and in GeV$^n$, respectively. Two options have been tested:

Option 1: $A$, $\langle r_p^2 \rangle$, $\langle r_p^4 \rangle$, $\langle r_p^6 \rangle$ are free parameters, $\langle r_p^8 \rangle$ is a fixed variable;

Option 2: $A$, $\langle r_p^2 \rangle$, $\langle r_p^4 \rangle$ are free parameters, $\langle r_p^6 \rangle$ and $\langle r_p^8 \rangle$ are fixed variables.

As it follows from this analysis, the $Q^8$ term plays unimportant role in determination of $R_p$. Therefore, one can fix $\langle r_p^8 \rangle$, for example, at the value $\langle r_p^8 \rangle = 374$ fm$^8$ from ref. [7] with uncertainty of the same order as the input parameter. But even such arbitrary large uncertainty introduces a systematic error in $R_p$ on a negligible level of $5 \cdot 10^{-4}$ fm. The statistical error in $R_p$ in the fits with four free parameters ($A$, $\langle r_p^2 \rangle$, $\langle r_p^4 \rangle$, $\langle r_p^6 \rangle$) is $\Delta R_p$ (stat) = ± 0.0085 fm with $\Delta R_p$ (syst) = ± 0.0005 fm. The statistical error can be reduced to $\Delta R_p$ (stat) = ± 0.0042 fm in the fit with three free parameters ($A$, $<r_p^2>$, $<r_p^4>$) by fixing $<r_p^6>$ to some value followed from analyses of the *ep* scattering data in the higher $Q^2$ region. However, in this case some systematic bias is introduced because of uncertainties in the $<r_p^6>$ value. As an example, taking $<r_p^6>$ = 30 (15) fm$^6$ from ref.[7], we obtain the systematic bias $\Delta R_p$ (syst) = ± 0.0025 fm.

**Conclusion**

The experimental method described in this report offers a new approach for measurements of the proton charge radius in the *ep* scattering experiments. The method is based on detection of the recoil protons in a hydrogen TPC operating as an "active target". The experiment is designed to provide high precision measurement of the *ep* elastic scattering differential cross section in the $Q^2$ range from 0.001 GeV$^2$ to 0.04 GeV$^2$ and to extract the proton charge radius with sub-percent precision. In comparison with the previous *ep* scattering experiments, the important advantages of the applied method are considerably lower radiative corrections, inherent to the recoil proton method, and absolute measurement of the differential cross sections. The experiment will be performed in the 720 MeV electron beam of the Mainz electron accelerator MAMI. The experimental setup is under construction with the plans to start measurements in 2020.

# ПРЕЦИЗИОННОЕ ИЗМЕРЕНИЕ ЗАРЯДОВОГО РАДИУСА ПРОТОНА В ЭЛЕКТРОН – ПРОТОННОМ РАССЕЯНИИ

## Проект

**А.А. Воробьев**

*Петербургский институт ядерной физики, НИЦ Курчатовский институт*

**Абстракт.** Представленный в докладе проект эксперимента основан на использовании нового метода исследования электрон-протонного рассеяния в области малых переданных импульсов. Основой метода является регистрация протонов отдачи с помощью водородной «активной мишени». Целью эксперимента является измерение сечения *ep* упругого рассеяния в области $Q^2$ от 0.001 ГэВ$^2$ до 0.04 ГэВ$^2$ с 0.1% относительной и 0.2% абсолютной точностью и определение величины протонного зарядового радиуса с суб-процентной точностью. В сравнении с предыдущими экспериментами по исследованию *ep* рассеяния, важным преимуществом данного эксперимента являются значительно меньшие радиационные поправки, присущие методу протонов отдачи, а также измерение абсолютного дифференциального сечения. Эксперимент будет проводиться с использованием 720 МэВ электронного пучка на электронном ускорителе MAMI в Майнце. Параметры этого пучка оптимальны для данного эксперимента. Проект эксперимента был одобрен Программным Комитетом MAMI и было подписано специальное Соглашение о проведении этого эксперимента между Петербургским институтом ядерной физики и Институтом ядерной физики Университета в Майнце. В этом проекте будут также принимать участие ученые из ГСИ Центра исследования тяжелых ионов, Германия; Объединенного института ядерных исследований, Россия; Колледжа Вильяма и Мэри, США; Маунт Аллисон Университета, Канада; Университета Регино, Канада; Сант Мэри Университета, США. Начало измерений намечено на 2020 год.